# New heat treatment to prepare high quality polycrystalline and single crystal MgB$_2$ in single process

### Mohamed H. Badr[a)] and K.-W. Ng
*Department of Physics and Astronomy, University of Kentucky, Lexington, KY 40506-0055, U.S.A.*

We report here on a new heat treatment to prepare both dense polycrystalline and single crystal MgB$_2$ high quality samples in one single process. Resistivity measurements for polycrystalline part of the sample gives a residual resistivity ratio RRR=16.6 and a very low normal state resistivity $\rho_o(40\ K) = 0.28\ \mu\Omega$cm. Both SEM and SQUID study on polycrystals reveal the high quality, dense character and well coupling of grain boundaries. On the other hand, the high quality single crystals have a unique shape that resembles the hexagonal crystal structure. SQUID measurements reveals very weak flux pinning character implying our single crystals to be very clean. In this study, we conclude that heat treatment is playing a major rule on the characteristics of both polycrystalline and single crystal MgB$_2$. Samples are thoroughly characterized by x-ray, resistivity, dc SQUID and SEM.

Since the discovery of superconductivity in MgB$_2$,[1] a great concern has been given to this material for both the interesting physics questions it raises and the industrial applications it promises. Superconductors put in practice so far are Nb47wt%Ti, Nb$_3$Sn, YBCO, and Bi-2223 with T$_c$'s of 9, 18, 92 and 108 K, respectively.[2] MgB$_2$ (T$_c$ = 39 K) can be a potential candidate in power applications for many reasons, like low-cost production of the basic materials and the ease of metalworking and fabrication. Moreover, unlike Bi-2223, grain boundaries in MgB$_2$ have a minimal effect on suppercurrent and enhance current density by pinning the magnetic flux inside it. In comparison to all practical superconductors, MgB$_2$ has the lowest normal state resistivity ($\rho_o(40\ K) < 1\ \mu\Omega$cm). Thus, MgB$_2$ magnet wires are expected to handle quenching more efficient than Nb47wt%Ti ($\rho_o(10\ K) = 60\ \mu\Omega$cm) and Nb$_3$Sn ($\rho_o(20\ K) = 5\ \mu\Omega$cm).

Reported properties on polycrystalline, thin film, and single crystal MgB$_2$ samples are widely varied depending on the required form and the procedure of preparation. Polycrystalline form of MgB$_2$ can be prepared by sealing a mixture of Mg and B with Mg:B = 1:2 in a tantalum tube, then heating at 950 °C for 2 h before quenching to room temperature.[3] MgB$_2$ single crystals are synthesized mainly by heat treatment in sealed metal containers[4] or by sintering at high temperature and pressure.[5] Single crystals can be obtained in sub-millimeter size, and their shapes are mostly irregular.[4] In general, both single crystal sizes and shapes seems to depend strongly on methods of preparation.

In this Letter, we report on a simple and single method to prepare high quality MgB$_2$ in both single crystal and polycrystalline forms in same treatment. The importance of this method is its simplicity where neither high pressure (GPa range) cells nor very high sintering temperatures (around 1700 °C) are required to prepare single crystals. Although our crystallites are in range of tens of microns, they have a unique shape closely resembles the hexagonal structure of MgB$_2$. Furthermore, this heat treatment produces high quality polycrystalline MgB$_2$ with well grain coupling, lowest reported normal state resistivity $\rho_o(40\ K) = 0.28\ \mu\Omega$cm, and high residual resistivity ratio of 16.6. Finally, preparing both single and polycrystalline MgB$_2$ under the same conditions gives a great opportunity to study the reported discrepancies in their transport and magnetic properties.

According to Naslain,[6] starting with the atomic ratio B/Mg =1.9 and heating at 1200-1400 °C assures an equilibrium between Mg vapor and liquid that generates an internal pressure. This pressure will be enough to form MgB$_2$ with the possibility of crystallization in the presence of small temperature gradient. Accordingly, starting with B/Mg = 1.9, about 2 gm of amorphous boron powder (99.99%, 325 mesh, Alfa Aesar) and Mg turnings (99.98%, 4 mesh, Alfa Aesar) are mechanically pressed in a tantalum tube (99.9%, 8.54 mm inner diam and 0.16 mm thick) at ambient pressure. The tantalum tube is then vacuum-sealed in a quartz tube and placed inside a box furnace in a nearly vertical position. The sample reached 1200 °C with heating rate of 700 °C/h and stayed there for 0.5 h, then cooled down to 1000 °C with a rate of 10 °C/h. Once the temperature reached 1000 °C, the cooling rate was further reduced to 2 °C/h until temperature reached 700 °C where furnace was turned off. The obtained batch consists of two separate portions. The first portion consists of hundreds of shiny separate single crystals, this will be denoted as S-sample. The base consists of one very dense polycrystalline piece with uniform golden-gray color. This portion will be donated as P-sample.

Crystal structure of P-sample is characterized by powder x-ray diffractometer (Scintag PAD V) with solid state detector and Cu K$_\alpha$ radiation. As seen in figure 1(a), all MgB$_2$ peaks are indexed and coincides with the standard diffraction pattern, figure 1(c). X-ray pattern shows no presence of un-reacted Mg or other Mg-B phases. However, two low intensity MgO peaks (Fig.1(a)) appear at 2θ = 42.9° and 62.4°. Existence of MgO may be due to starting with excess Mg and sealing Ta tube at ambient pressure. Resistivity measurements are performed by the conventional four-probe technique. Figure 2 shows the temperature-dependence of resistivity ρ(T) for P-sample at a constant current of 53 mA and zero-field. As seen, P-sample has a transition temperature T$_c$ = 38.4 K (2 % criteria) with sharp transition width ΔT$_c$ = 0.2 K (10 % - 90 % criteria).

---
[a)]Author for whom correspondence should be addressed. Also on leave from: Department of Physics, Faculty of Science, Menoufiya University, Shebin El-Koom 32511, Egypt. Electronic mail: mhbadr0@uky.edu

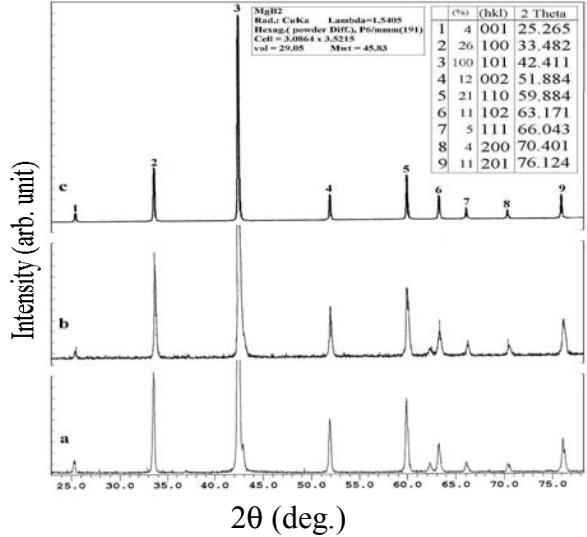

Fig. 1. X-ray diffraction pattern for (a) P-sample, (b) polycrystalline $MgB_2$ studied previously [Ref. 11], and (c) powder $MgB_2$ from standard database, full characteristics are given in the inserted tables.

Furthermore, the sample has RRR = $(R_{300K}/R_{40K})$ = 16.6 and $\rho_o$ (40 K) = 0.28 $\mu\Omega$cm. To our best knowledge, this is the lowest normal state resistivity ever reported for $MgB_2$. These features promise $MgB_2$ magnet wires to handle quenching very efficiently.

While $MgB_2$ single crystals have RRR ≈ 5,[4,7] polycrystalline samples have reported values of RRR > 20.[8] Jung et al.[9] have referred the observed high RRR (as compared to RRR of single crystals) in polycrystalline $MgB_2$ to presence of Mg impurities with its very large RRR. Accordingly, They considered such high RRR of polycrystalline $MgB_2$ as an extrinsic property. On the other hand, Ribeiro et al.[10] have studied the effects of boron purity, boron isotope, and Mg content on RRR of polycrystalline $MgB_2$. In that paper, they have considered using boron isotope $^{11}B$ (RRR ≈ 18 for $Mg^{11}B_2$) rather than natural boron B (RRR ≈ 7 for $MgB_2$) as the main key to achieve high quality samples with high RRR. Furthermore, the observed high RRR in their polycrystalline $Mg^{11}B_2$ has been accounted as an intrinsic property. To clarify such discrepancy about the nature and rule to achieve high RRR in polycrystalline $MgB_2$, it will be noteworthy to share our experience in this issue. In a previous article,[11] we have prepared polycrystalline $MgB_2$ following Bud'ko's et al. procedure.[3] In that paper, our sample has RRR = 8 close to RRR ≈ 7 of polycrystalline $MgB_2$ prepared under similar conditions and with same boron powder (99.99%, 325 mesh, Alfa Aesar).[10] Figure 1(b) shows x-ray spectra for that sample. As can be deduced by comparing the two charts (Figs. 1(a) and 1(b)), both samples almost have same MgO content and both show no noticeable trace of un-reacted Mg or other Mg-B phases. The main difference between our previously prepared sample (RRR = 8) and the P-sample (RRR = 16.6) is heat treatment. Therefore, under no circumstances the high RRR of P-sample could be attributed to Mg impurities. On the other hand, although we agree with Ribeiro et al.[10] in considering high RRR as an intrinsic property, our interpretation is different. It is clear from our argument mentioned above that using same starting Mg and natural B the residual resistivity ratio can be varied for different heat treatments. For that, we believe both $\rho_o$ and RRR can be enhanced by enhancing the coupling of grains boundaries. Our new heat treatment has greatly enhanced the coupling (as SEM will confirm below) resulting in the very low $\rho_o$(40 K) = 0.28 $\mu\Omega$cm and RRR as high as 16.6.

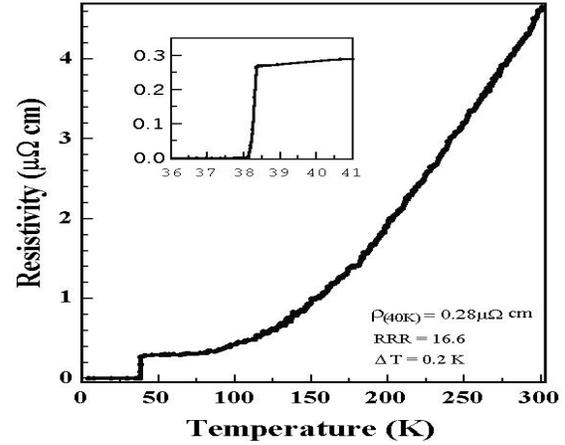

Fig. 2. Resistivity vs temperature for polycrystalline $MgB_2$ (P-sample). The insert is a close up of the transition region.

To further characterize P-sample, Fig. 3(a) shows scanning electron microscope (SEM) picture of the sample's surfaces. Surface morphology reflects the polycrystalline, dense character, and well coupling of grains (no clear boundaries can be observed) as compared to previously reported SEM pictures.[12]

Although sub-millimeter single crystals have the advantage of size, they have many disadvantages resulted mainly from their preparation under high pressure and temperature. For instance, in $MgB_2$ single crystals grown under high pressure, a deficiency in Mg (4%) has been observed.[13] Furthermore, the high temperature gradient used to grow sub-millimeter crystals causes growing instabilities that resulted in irregularly shaped crystals.[14] Therefore, we consider the reported discrepancies in transport and magnetic properties of sub-millimeter single crystals to be structurally-related, and optimizing their growing techniques still a challenge. Figure 3(b) shows SEM picture for S-sample where superconductivity in these single crystals has been proven by magnetization measurements as will be discussed below. SEM picture shows single crystals to have an average diagonal length of 50μm and thickness of about 10 μm. The angles formed by the surfaces reveals the hexagonal structure of $MgB_2$ crystals. In comparison with the previously reported $MgB_2$ single crystals[15] and to our best knowledge, our single crystals are unique in shape. Both the size and shape can be attributed to the expected low growing rate due to both low sintering temperature and small temperature gradient.

A dc superconducting interface device (SQUID, Quantum Design MPMS) magnetometer was used to measure temperature–dependent magnetization, M(T). Figure 4 (left scale) gives M(T) at low field of 100 Oe for P-sample with $T_c$ = 39.1 K (2 %) and transition with $\Delta T_c$ = 0.7 K (10% - 90%) indicating bulk superconductivity of the sample. M(H) measurements predicts P-sample to have a lower critical field $H_{c1}$ ≈ 0.1 T at T= 25 K, figure not included. Field-cooled (FC) mode gives a very small magnetization signal which is less than 0.4 % of zero–field-cooled (ZFC) signal. This can be attributed to large flux trapping in the well coupled grain boundaries. The well coupling of these boundaries is also evidenced by SEM picture (Fig. 3a) and the low normal state resistivity, figure 2. Such properties put polycrystalline $MgB_2$ as a potential candidate in high current applications.

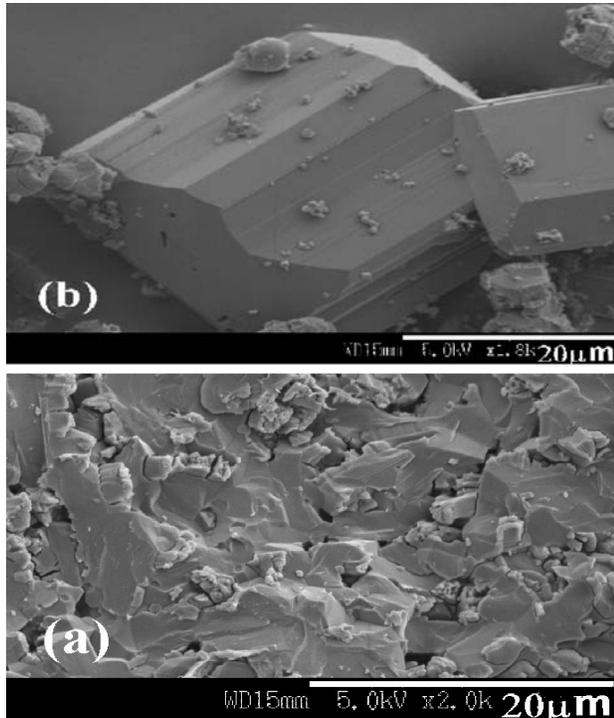

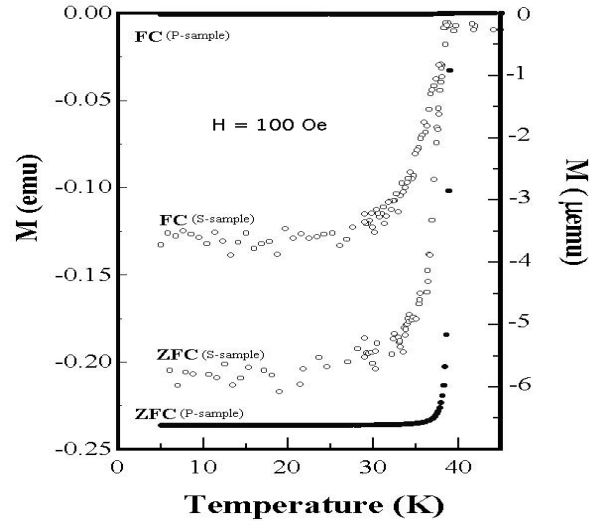

Fig. 3. SEM pictures for (a) polycrystalline $MgB_2$ (P-sample) and (b) single crystal $MgB_2$ (S-sample).

To measure M(T) for S-sample, we collected about 200 randomly oriented crystallites on a non-magnetic strip. Figure 4 (right scale) shows M(T) for both ZFC and FC modes at low field of 100 Oe. ZFC mode reveals a superconducting transition with width $\Delta T_c$ = 4.6 K (10%-90%) and $T_c$ = 38.5 K (2 %). This transition width is much less than that reported by other groups. For instance, Xu et al.[4] prepared $MgB_2$ single crystals in mm-range with estimated transition width $\Delta T_c \approx$ 17 and 10 K (10% - 90%) for magnetic fields parallel and perpendicular to boron planes, respectively. The observed broad transition in S-sample (compared to P-sample) could be attributed to the randomly oriented character of S-sample. This view is supported by a study of Eltsev et al.[16] in which they have found the transition width to depend on the direction of magnetic field relative to boron plane. Figure 4 shows the FC signal to be about 65 % of ZFC signal. Such large FC signal reflects the very weak flux pinning character of single crystals. This requires our single crystals to be very clean, i.e. free from flux trapping centers.

In summary, we have prepared both polycrystalline and single crystal $MgB_2$ using a new heat treatment. Polycrystalline part has an RRR = 16.6 and $\rho_o(40 K)$ = 0.28 $\mu\Omega$cm. SEM and SQUID study reveal the dense character, high quality, and well coupling of grains. On the other hand, single crystals have high quality with unique shape that resembles the hexagonal crystal structure.

Fig.4. Temperature-dependent magnetization curves for polycrystalline (left scale) and single crystal $MgB_2$ (right scale).

This work was supported by NSF Grant No. DMR9972071. The corresponding author would like to thank the Egyptian Ministry of Higher Education & Scientific Research for supporting this work as well.

[1] J. Nagamatsu, N. Nakagawa, T. Muranaka, Y. Zenitani, and J. Akimitsu, Nature (London) **410,** 63 (2001).

[2] D. Larbalestier, A. Gurevich, D. M. Feldmann, and A. Polyanskii, Nature (London) **414,** 368 (2001).

[3] S. L. Bud'ko, G. Lapertot, C. Petrovic, C. E. Cunningham, N. Anderson, and P. C. Canfield, Phys. Rev. Lett. **86,** 1877 (2001).

[4] See for example: M. Xu, H. Kitazawa, Y. Takano, J. Ye, K. Nishida, H. Abe, A. Matsushita, N. Tsujii, and G. Kido, Appl. Phys. Lett. **79,** 2779 (2001).

[5] See for example: M. Angst, R. Puzniak, A. Wisniewski, J. Jun, S. M. Kazakov, J. Karpinski, J. Roos, and H. Keller, Phys. Rev. Lett. **88,** 167004 (2002).

[6] R. Naslain, A. Guette, and M. Barret, J. Solid State Chem. **8,** 68 (1973).

[7] See for example: S. Lee, H. Mori, T. Masui, Y. Eltsev, A. Yamamoto, and S. Tajima, eprint cond-mat/0105545.

[8] P. C. Canfield, D. K. Finnemore, S. L. Bud'ko, J. E. Ostenson, G. Lapertot, C. E. Cunningham, and C. Petrovic, Phys. Rev. Lett. **86,** (2001).

[9] C. U. Jung, H.-J. Kim, M.-S. Park, M.-S. Kim, J. Y. Kim, Z. Du, S.-I. Lee, K. H. Kim, J. B. Betts, M. Jaime, A. H. Lacerda, and G. S. Boebinger, eprint cond-mat/0206518.

[10] R. A. Ribeiro, S. L. Bud'ko, C. Petrovic, and P. C. Canfield, eprint cond-mat/0204510.

[11] M. H. Badr, M. Freamat, Y. Sushko, and K. W. Ng, Phys. Rev. B **65,** 184516 (2002).

[12] See for example: J.-S. Rhyee, C. A. Kim, B. K. Cho, and J.-T. Kim, Appl. Phys. Lett. **80,** 4407 (2002).

[13] S. L. H. Mori, A. Yamamoto, S. Tajima, S. Sato, Phys. Rev. B **65,** 095207 (2002).

[14] S. Lee, T. Masaui, H. Mori, Y. Eltsev, A. Yamamoto, and S. Tajima, eprint cond-mat/0207247.

[15] See for example: Y. Machida, S. Sasaki, H. Fujii, M. Furuyama, I. Kakeya, and K. Kadowaki, eprint cond-mat/0207658.

[16] Y. Eltsev, S. Lee, K. Nakao, N. Chikumoto, S. Tajima, N. Koshizuka, and M. Murakami, eprint cond-mat/0202133.